\documentclass[useAMS,usenatbib]{mn2e}
\usepackage{graphicx}
\usepackage{latexsym}
\usepackage{xspace}
\usepackage{fixltx2e}
\usepackage{pslatex}

\title[Accretion Luminosity and Primordial Fragmentation]{The Effects of Accretion Luminosity upon Fragmentation in the Early Universe}
\author[Smith et~al.]{Rowan J. Smith$^{1}\thanks{Email: rowan@ita.uni-heidelberg.de}$, Simon C. O. Glover$^{1}$, Paul C. Clark$^{1}$ \and Thomas Greif$^{2}$, Ralf S. Klessen$^{1}$ \\
$^1$ Zentrum f\"ur Astronomie der Universit\"at Heidelberg, Institut f\"ur Theoretische Astrophysik, Albert-Ueberle-Str. 2, 69120 Heidelberg, Germany \\
$^2$ Max-Planck-Institut f\"ur Astrophysik, Karl-Schwarzschild-Stra§e 1, 85740 Garching bei M\"unchen, Germany
 }

\begin{document}

\pagerange{\pageref{firstpage}--\pageref{lastpage}} \pubyear{2010}

\maketitle

\label{firstpage}

\def\mnras{MNRAS}
\def\apj{ApJ}
\def\aap{A\&A}
\def\apjl{ApJL}
\def\apjs{ApJS}
\def\bain{BAIN}
\def\pasp{PASP}
\def\araa{ARA\&A}
\def\ga{\sim}
\def\nat{Nature}
\def\aj{AJ}
\def\pasj{PASJ}
\def\na{New Astronomy}
\def\nar{New Astronomy Reviews}

%some simplifying definitions

\newcommand{\eq}{Equation }
\newcommand{\fig}{Figure }
\newcommand{\tab}{Table }

\newcommand{\msun}{\,M$_{\odot}$\,}
\newcommand{\rsun}{\,R$_{\odot}$}
\newcommand{\lsun}{\,L$_{\odot}$}
\newcommand{\myr}{\,M$_{\odot}$\,yr$^{-1}$}
\newcommand{\h}{H$_2$\xspace}

\newcommand{\gcmc}{\,g\,cm$^{-3}$\xspace}
\newcommand{\cmc}{\,cm$^{-3}$\xspace}
\newcommand{\gcms}{\,g\,cm$^{-2}$\xspace}
\newcommand{\kms}{\,kms$^{-1}$\xspace}

\newcommand{\E}{\times 10}

\begin{abstract}
We introduce a prescription for the luminosity from accreting protostars into smoothed particle hydrodynamics simulation, and apply the method to simulations of five primordial minihalos generated from cosmological initial conditions. We find that accretion luminosity delays fragmentation within the halos, but does not prevent it. In halos that slowly form a low number of protostars, the accretion luminosity can reduce the number of fragments that are formed before the protostars start ionising their surroundings. However, halos that rapidly form many protostars become dominated by dynamical processes, and the effect of accretion luminosity becomes negligible. Generally the fragmentation found in the halos is highly dependent on the initial conditions. Accretion luminosity does not substantially affect the accretion rates experienced by the protostars, and is far less important than dynamical interactions, which can lead to ejections that effectively terminate the accretion. We find that the accretion rates onto the inner regions of the disks (20~AU) around the protostars are highly variable, in contrast to the constant or smoothly decreasing accretion rates currently used in models of the pre-main sequence evolution of Population III stars.
\end{abstract}


\section{Introduction}
Over the last decade the picture of star formation in the primordial universe was that of single massive stars that were the sole inhabitants of the first dark matter minihalos formed after the big bang. This was the inference made from the numerical models of \citet{Abel00,Abel02} which was later reinforced by the findings of \citet{Yoshida08}. In these works the star formation process was followed from cosmological initial conditions, right up to the densities where the first protostar formed. This required the combination of cutting-edge hydrodynamic simulations with a detailed treatment of the chemistry and thermodynamics of the gas, as this controls where fragmentation occurs \citep[e.g.][]{Dalgarno87,Abel97,Galli98,Glover01,Omukai05}. From this \citet{Omukai03} used detailed stellar modelling to estimate a final mass of the primordial star and found that above an accretion rate of $4\E^{-3}$\myr masses in excess of $100$ \msun were produced.

However the above mentioned simulations were unable to proceed beyond the first stage of collapse due to the numerical difficulty of following the evolution of high density gas, where the numerical time step becomes prohibitively small.  More recent work, which follows the collapse beyond the formation of the first protostellar core, has cast doubt on the isolated picture of Population III star formation, suggesting that they may have been members of binaries, multiples, or even small-N clusters. Using an idealised Bonnor-Ebert model, \citet{Machida08} showed that binary stars were the likely outcome of a rotating minihalo. \citet{Clark08} showed that in idealised conditions, gas with a barotropic equation of state approximating the behaviour of primordial gas could fragment vigorously to form a small cluster. This work took advantage of the `sink' particle technique used in present day star formation \citep{Bate95} to follow the evolution past the formation of the first object by replacing high density collapsing gas with a non-gaseous particle that can accrete additional bound gas but that only interacts with its surroundings via gravity. A follow up study to this work \citep[][a]{Clark11a} found that the fragmentation persisted when the detailed chemical and thermodynamic evolution of the gas was followed.

Although these simulations make use of idealised initial conditions, other work has addressed this issue with cosmological initial conditions and full chemical networks. \citet{Turk09} using the AMR (adaptive mesh refinement) method showed that a wide binary with a separation of $800$~AU formed in one out of the five minihalos that they simulated. \citet{Stacy10} used smoothed particle hydrodynamics (SPH) combined with the sink particle technique to show that there is further fragmentation after the first object forms in a minihalo and that a small multiple system can be formed. \citet[][b]{Clark11b} found that the individual disks around Population III stars are prone to fragmentation and are likely to fragment into higher-order multiples. This result was confirmed more recently by \citet{Greif11} using the novel moving mesh AREPO code \citep{Springel10}. 

The case for fragmentation therefore seems robust, as it has been found by multiple authors using complementary numerical techniques. If this result withstands further investigation, then it will have important implications for our understanding of cosmology and the early universe. For instance, the first stars are thought to be an important source of ionizing photons in the early universe, and hence an important contributor to the reionisation of the intergalactic medium \citep{Kitayama04,Sokasian04,Whalen04,Alvarez06,Johnson07}. If the mass available for accretion is split into multiple stars there may be fewer truly massive stars, which would reduce the number of ionizing photons produced. Likewise, the eventual fate of a Population III star, and consequently its enrichment of the surrounding gas, is determined by its mass \citep{Fryer01,Heger03,Yoshida04}. Additionally, while the traditional picture of extremely massive single stars meant there would be few observational signatures that would be observable today, the new picture of a few multiple stars leads to some new observational possibilities. In the simulations of \citet{Greif11} there are many dynamical interactions between the protostars that lead to the ejection of some of the low-mass protostars. There is therefore the exciting possibility that these stars could still be shining today, providing a direct insight into the physical conditions in the high-redshift universe. Moreover the possibility of tight binary systems resulting from disk fragmentation \citep[][b]{Clark11b} would allow primordial X-ray binaries, cataclysmic variables, or even gamma ray bursts. 

Given the potential implications of these results, it is vital to consider whether there are any mechanisms that could suppress fragmentation within primordial minihalos. Ionisation from the protostar is the most likely mechanism to stop further fragmentation. \citet{Omukai02} showed that above a critical ionising flux an HII region from a primordial star can unbind the surrounding gas, while ionising fluxes below this value may nevertheless suppress fragmentation. \citet{Tan04} use a semi-analytic model to describe the evolution of feedback from a steadily growing protostar and find that once the star is older than its Kelvin-Helmholtz time, and is contracting towards the main sequence, there is a rapid increase in the amount of ionising radiation it emits. One could speculate that this transition may therefore mark the point at which fragmentation is halted. In simulations of local star formation, ionisation has been found to be unable to prevent fragmentation \citep[][a,b,c]{Peters10}. However, it is likely that the effect of ionising radiation was more significant in the early universe, since it would have been accompanied by  the dissociation of \h by Lyman-Werner photons, thereby removing the primary gas coolant \citep{Omukai99,Glover01}.

The early period of protostellar growth before the ionising radiation becomes important therefore represents the most favourable window of opportunity in which fragmentation can occur. One of the few processes that would act against fragmentation in this epoch is accretion luminosity feedback from the forming protostars. We first introduced this in the simulations reported on in \citet[][b]{Clark11b} in order to study its effects on the stability of Pop.\ III accretion disks. We found that accretion luminosity does indeed change the disk evolution, but that it cannot ultimately prevent the disk from fragmenting. The feedback is able to support the inner $20$ AU of the disk, which was previously unstable, against fragmentation, but the outer regions still fragment, albeit after a longer time period. A more detailed
description of how fragmentation in the disk proceeds can be found in \citet[][b]{Clark11b}.

In this work we seek to address a different question. The optimal time for fragmentation within a minihalo is the first few thousand years before the first protostar approaches the main sequence, at which point ionisation feedback will become significant and will act against further fragmentation. In this work we aim to capture the full evolution of the halo during this regime to determine how much fragmentation can occur in this time interval, and to what extent radiation affects the fragmentation.

\section{The Method}\label{method}
We perform the calculations for this paper using the SPH code GADGET 2 \citep{Springel05}. We have substantially modified this code to include a fully time-dependent chemical network, details of which can be found in the appendix of \citet[][a]{Clark11a}. Our treatment includes: \h  cooling \citep{Glover08}; optically thick \h cooling using the Sobolov approximation \citep{Yoshida06}; collisionally induced \h emission \citep{Ripamonti04}; ionisation and recombination \citep{Glover07a}; as well as heating and cooling from changes in the chemical makeup of the gas, and heating and cooling from shocks, compression and expansion of the gas. \citet{Turk11} showed that the choice of \h three-body formation rate coefficient influences the resulting dynamics of the gas within the halo. In this work we use the three-body \h formation rate of \citet{Glover08b} which is intermediate within the range of the published rates and is based on applying the principle of detailed balance to a relatively recent calculation of the collisional dissociation of \h.

We include heating from the accretion luminosity as an additional heating term when solving the ordinary differential equations that govern the chemical and thermal evolution of the gas. While the protostar will also have an interior luminosity, for the majority of the early protostellar evolution this is an order of magnitude lower than the accretion luminosity \citep{Hosokawa09}, and so we focus here on only the accretion luminosity. The accretion luminosity will also transfer some momentum to the gas. However, the resulting outward force is many orders of magnitude smaller than that of the gravitational force on the gas, and so it is safe to neglect it.

The accretion luminosity is calculated from the standard equation,
\begin{equation}
L_{acc}=\frac{GM_*\dot{M}}{R_*}
\end{equation} 
where $\dot{M}$ is the accretion rate of the protostar and $R_*$ is the stellar radius. We make the assumption that the gas is optically thin to the emitted radiation which ensures that we are overestimating the effects of the accretion luminosity to obtain an upper limit on the feedback effects. The heating rate per unit mass for the gas will then become
\begin{equation}
 \begin{array}{ll}
    \Gamma_{acc}=\rho_g \kappa_P \left( \frac{L_{acc}}{4\pi r^2} \right) & \mbox{erg  g$^{-1}$ s$^{-1}$} \\
 \end{array}
\end{equation}
where $\rho_g$ is the gas density,  $\kappa_P$ is the Plank mean opacity and $r$ is the distance of the gas from the source. We calculate the Planck mean opacity of the gas by interpolating from the tables of \citet{Mayer05} which include contributions from deuterium and lithium in the gas in addition to hydrogen and helium.

To accurately calculate the accretion luminosity, both the accretion rate and the stellar radius need to be known. We achieve this by using sink particles \citep{Bate95} to model the protostars and record their growth in mass throughout the simulation. These were first implemented into Gadget by \citet{Jappsen05}. Sink particles are non-gaseous particles that replace extremely dense gas if it is found to be both gravitationally bound and collapsing. This allows us to evolve of the simulation beyond the point at which the first object forms. For a recent discussion of the pro's and cons of sink particles see \citet{Federrath10a}. We form sink particles above densities of $10^{15}$\cmc, after which there are no more chemical heating terms that can prevent the gas collapsing to form a protostar. The sinks have accretion radii of $20$ AU and consequently the inner accretion disk is not resolved (although any disk outside this distance is resolved). \citet[][b]{Clark11b} and \citet{Greif11} consider fragmentation within this regime and the effect of accretion luminosity upon it. In this work, we use larger sink radii to allow us to study the cluster as a whole.

The accretion rate of the sinks can be found from simply looking at how their mass grows in time. However, as SPH is a particle-based method, accretion occurs in discrete units and can be noisy. In order to account for this, we calculate the accretion rate by taking a smoothed average over the last 
100 years, updated at 10 year intervals. For the typical accretion rates of the sinks in our simulation this equates to between 0.1 and 1\msun of accreted material. This is equivalent to the accretion rate being smoothed over $10^3 - 10^4$ particles (and updated only after at least 100 new particles have been added) and therefore we can be sure that any variability in the accretion rates is not due to particle noise. 

In reality, material will flow on to the protostar through the inner disk, which will delay it from reaching the protostar. Inwards mass transport typically takes place over the viscous timescale, which for a thin disk is typically much larger than the dynamical timescale \citep{Pringle81}. However in the primordial protostellar disk study of \citet[][b]{Clark11b}, the disk had a scale height of approximately 5 AU which, given that we are considering an inner disk of 20 AU, means that the thin disk approximation is not valid. Furthermore, the disk was self-gravitating and was rapidly transferring its angular momentum through gravitational torques which lead to high accretion rates. Given these findings, as we do not resolve this region, we simply update the accretion rate immediately after it is calculated. However, the procedure of averaging the accretion rates which we adopt for numerical reasons will to some extent mimic the effect of accreted material being buffered by the inner disk.

Accurately finding the stellar radii is a complex problem that would require the implementation of detailed stellar evolution models within our hydrodynamic simulation which is beyond current computational resources. Instead we used the models of \citet{Omukai03} to derive a simple power law approximation of the stellar radius. In the early stages of the protostellar evolution the cooling time of the interior is very long and the protostar evolves adiabatically. \citet{Stahler86a} showed that the stellar radius during this phase grows according to
\begin{equation}
R_{*}=26M_*^{0.27}\left(\frac{\dot{M}}{10^{-3}}\right)^{0.41} \mbox{	\rsun}
\end{equation} 
where $R_*$ is the stellar radius in \rsun \xspace, $M_*$ is the current stellar mass in \msun and $\dot{M}$ is the accretion rate in \myr. After some time, the internal heat becomes sufficient  to drive an outward luminosity wave, which results in the rapid swelling of the stellar radius. Once the luminosity wave reaches the stellar surface, the interior achieves a relaxed state and undergoes Kelvin-Helmholtz contraction until the radius shrinks to its main sequence value. 

\begin{figure}
\begin{center}
\includegraphics[width=3in]{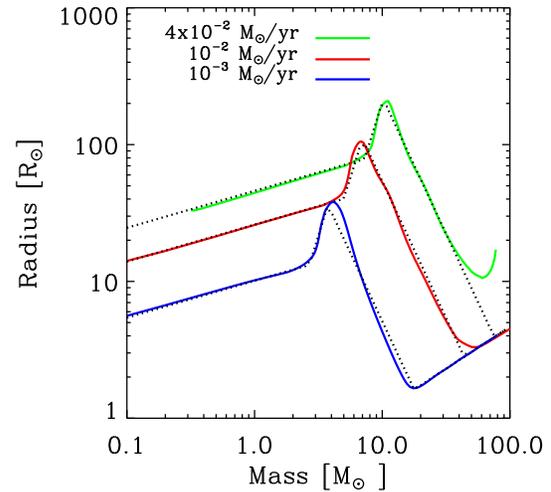}
\caption{The stellar radius as a function of mass found in the stellar evolution models of \citet{Omukai03} for three different accretion rates. The dotted black lines show the stellar radius given by our semi-analytic model for these accretion rates.}
\label{radmodel}
\end{center}
\end{figure}

\fig \ref{radmodel} shows the variation of stellar radius with mass calculated by \citet{Omukai03}. Using the model with an accretion rate of $10^{-3}$\myr as our fiducial case, we found that the stellar radius for this accretion rate could be described by,
\begin{equation}\label{imfeq}
R_* \propto \left\{
    \begin{array}{ll}
         26 M_*^{0.27}(\dot{M}/10^{-3})^{0.41} &  M_*\leq p_1 \\
         A_1 M_*^{3} & p_1 \leq M_* <p_2 \\
         A _2 M_*^{-2} & p_2 \leq M_* \mbox{ \& } R< R_{ms} \\
    \end{array}
\right.
\end{equation}
where $p_1=5$\msun is the transition between the adiabatic phase and the luminosity wave, and $p_2=7$\msun is the transition between the luminosity wave phase and the Kelvin-Helmholtz stage. The constants $A_1$ and $A_2$ are determined at these transition zones to give a continuous function. At $R_{ms}$ the radius has shrunk to its main sequence value and the feedback will be totally dominated by ionising radiation. The main sequence radius in \rsun \xspace for these stars from \citet{Omukai03} is 
\begin{equation}
R_{ms}=0.28 M_*^{0.61}  \mbox{	\rsun}
\end{equation}
To generalize this prescription to the case where $ \dot{M} \neq 10^{-3}$ \myr, one must account for the
fact that the transition points $p_1$ and $p_2$ between the phases scale with the accretion rate as
\begin{equation}
    \begin{array}{l}
	p_1=5 \dot{M}^{0.27}  \mbox{	\msun}, \\
	p_2=7 \dot{M}^{0.27} \mbox{	\msun}. 
     \end{array}
\end{equation}
The constants $A_1$ and $A_2$ must be correspondingly adjusted to maintain a smooth function. The resulting model is over-plotted onto \fig \ref{radmodel} and can be seen to qualitatively capture the change in stellar radius with mass. An important caveat here is that the models of \citet{Omukai03} are calculated using a constant accretion rate, which we shall later to show to be a poor assumption. However, the current model is a first step that couples the accretion rates, masses and stellar radii in real time with both the dynamics and the effects of the protostar's own radiation. 

\begin{figure*}
\begin{center}
\includegraphics[width=4.5in]{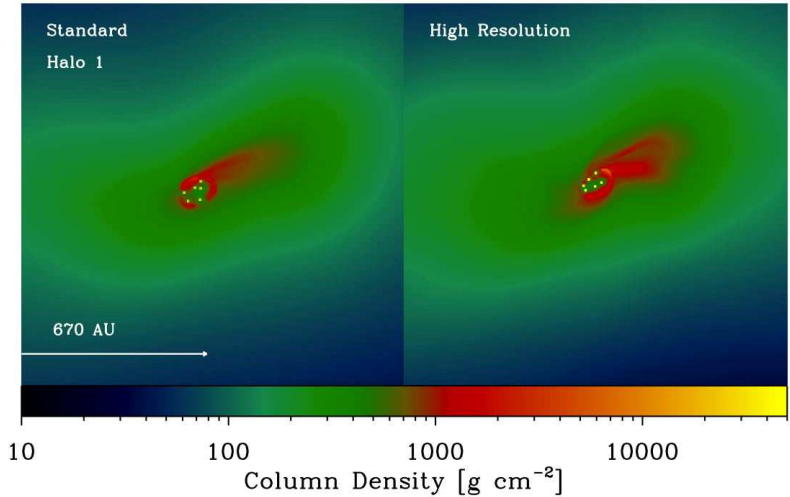}
\caption{Column density projection of the centre of Halo 1 just after the first burst of fragmentation simulated at our standard resolution and at ten times higher resolution. Feedback is present in both cases. The same number of fragments form in each case and in similar places.}
\label{resolution}
\end{center}
\end{figure*}

\section{Initial conditions}

In order to accurately capture the true properties of the first minihalos within which primordial stars form, we use the cosmological simulations of \citet{Greif11} as initial conditions. \citet{Greif11} presented simulations of primordial minihalos that strongly fragmented. These simulations made use of the novel moving mesh code AREPO \citep{Springel10} to fully resolve the formation of five minihalos from truly cosmological simulations.  Cells were refined during the evolution to ensure that the Jeans length was always resolved by at least 128 mesh points. The refinement was deactivated at densities greater than $n_H=10^9$ \cmc by which point the mass of each element was around $10^{-4}$ \msun. All of the halos modelled by \citet{Greif11} form multiple protostars with a range of masses. Some protostars are even ejected and stop accreting entirely. These minihalos, therefore, represent the perfect sample in which to test the effect of accretion luminosity upon fragmentation and to compare the magnitude of any effects to those of cosmic variance and dynamical interactions.

For this work we cut out the central two parsecs of the \citeauthor{Greif11} simulations and continue their evolution using our modified version of Gadget 2 with feedback as discussed in the previous section. Each mesh point in AREPO is converted to an SPH particle with the same properties as the original element. As the chemical network that is implemented in AREPO is based on the network we use here \citep[][b]{Clark11b} the chemical abundances can also be transferred from the original cosmological simulation. All the fragmentation and accretion takes place in the central region of the halos where the SPH particle masses are $10^{-4}$ \msun which gives us a mass resolution of at least $10^{-2}$ \msun \citep{Bate97}.

\begin{table}
	\centering
	\caption{The initial state of the inner 2~pc of the five minihalos. From each minihalo two simulations were run, one with feedback, and a control run with no feedback. See \citet{Greif11} for a more detailed description of the halos.}
		\begin{tabular}{c c c c c c c }
   	         \hline
	         \hline
	         Name & M [\msun] & $\bar{n}$ [\cmc] & $\bar{T}$ [K] \\
	         \hline
	         mh1  & 1810 & $4.62 \E^7$ & 409 \\
	         mh2  & 1240 & $8.09\E^7$ & 329 \\
	         mh3  & 1030 & $6.31\E^7$ & 292 \\
	         mh4  & 2000 & $7.92\E^7$ & 458 \\
	         mh5  & 3340 & $8.60\E^7$ & 494 \\
		\hline
		\end{tabular}
	\label{table-sim}
\end{table}

\tab \ref{table-sim} shows the initial conditions of the five halos at the point where our simulations commenced. Each halo is simulated twice, once with and once without feedback as a reference case. The halos typically contain a few thousand solar masses, with the largest having 3000 \msun. The mean densities are over $10^7$ \cmc and the mean temperature are around $300 - 500$ K. The gas is primarily atomic with most of the molecular hydrogen being contained within a disk-like region at the centre of the halos. A more detailed discussion of the physical properties of the considered halos is given in \citet{Greif11}.

As a resolution study we also increased the resolution of mh1 by a factor of ten using particle splitting \citep{Kitsionas02}. \fig \ref{resolution} shows a column density projection of the centre of halo 1 with feedback in our standard run and in the increased resolution run. In the higher resolution run smaller scale structure is resolved, but the same number of sinks are formed in both cases. The increased resolution decreases the numerical viscosity of the simulation which allows the disk to form earlier in the simulation as it drains more slowly due to the reduced angular momentum transport. Consequently the high resolution image of \fig \ref{resolution} is shown at an earlier snapshot in the simulation than the low resolution case. However the fragmentation of the disk is qualitatively the same once it forms in both cases. The relatively large accretion radius of our sink particles ($\sim$ 20AU) means that all structure smaller than this scale is swallowed by the sink and this limits the differences between the high and low resolution runs.

\section{Results}
\subsection{Fragmentation}

\begin{figure*}
\begin{center}
\includegraphics[width=6.5in]{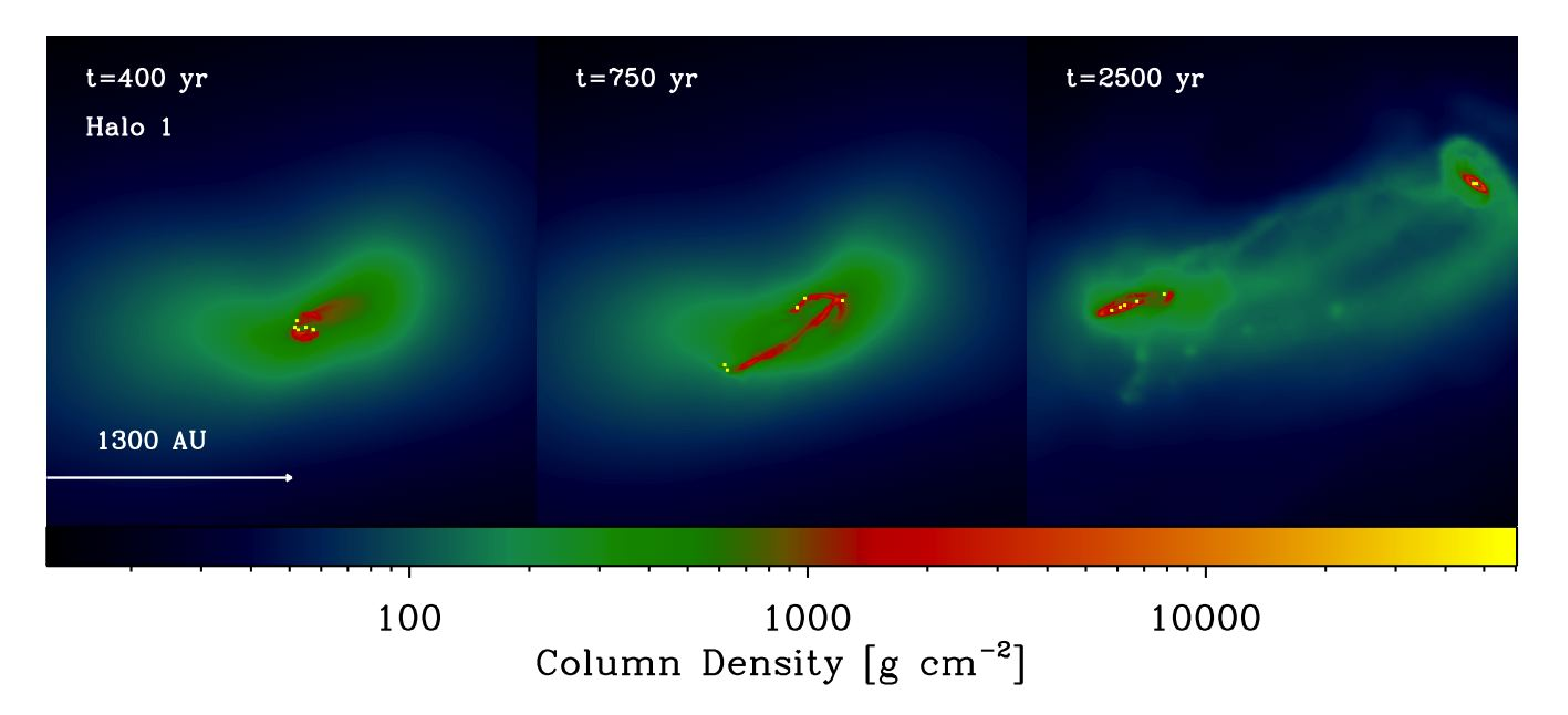}\\
\includegraphics[width=6.5in]{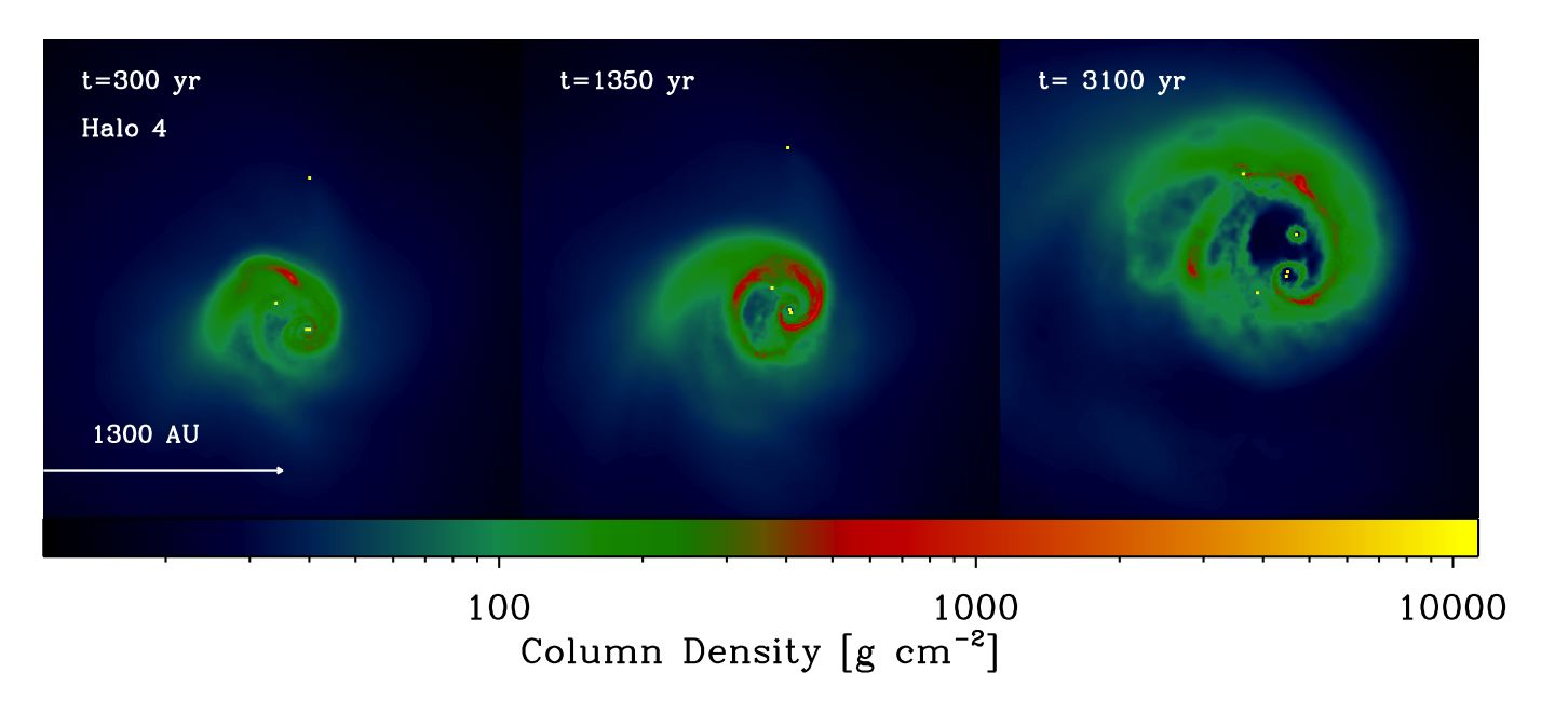}\\
\caption{Column density projection of fragmentation seen in the central 1300AU in the first few thousand years for halos mh1 (top) and mh4 (bottom). Sink particles are denoted by yellow squares. Despite the presence of accretion luminosity heating there is still fragmentation.}
\label{column}
\end{center}
\end{figure*}

\fig \ref{column} shows a column density projection of the central regions of halos 1 and 4 which are good examples of the fragmentation seen in all of the halos. In each case, a disk-like structure is formed due to the inability of the halo to transfer angular momentum outwards quickly enough during collapse. While the central region is disk-like, it is more extended and irregular than a classical disk. In every case this region fragments. Generally several fragments form almost simultaneously as the conditions for fragmentation are reached at multiple locations within the disk. Halos 1 and 5 fragment vigorously, whereas halos 2 and 3 fragment more slowly. Halo 4 is the case that is most affected by accretion luminosity and has a rate of fragmentation intermediate between the other cases.

Let us now consider whether the fragmentation seen here is resolved. While the central regions of our halos are disk-like, they are strongly self gravitating and have a large vertical extent. As an example let us consider the central regions of halo 4 which bears the greatest resemblance to a classical disk out of the set of simulations. Taking a density cut of material greater than $10^{10}$ \cmc the disk-like region had a radius which varies between 200 to 300 AU, and has a vertical extent of 150 AU measured from the mid plane. The thin disk approximation requires that the vertical extent of the disk, $H$, is much smaller than the radial extent of the disk, $R$, and is only appropriate for disks where $H/R \approx 0.1$ or lower \citep{Lodato08}.  \citet{Nelson06} propose that to avoid artificial fragmentation in disks the scale height must be resolved by 4 SPH particle smoothing lengths per scale height. In the inner 100 AU of the disk in halo 4, the average particle smoothing length is 14.3 AU. However, it is unclear how best to define a scale height for this irregular puffy disk. Given that even our best-case scenario cannot be considered a classical disk, and that as the simulations proceed fragmentation increasingly takes place in large spiral arm features or filaments (as seen in the later panels of \fig \ref{column}), a better resolution criteria to use is the Jeans mass. At our sink creation density of $10^{15}$ \cmc, the Jeans mass is $2\E^{-2}$ \msun, at least two times more massive than our resolution limit. Further to this, as shown in \fig \ref{resolution}, even when our resolution was increased by a factor of ten, the same number of fragments were formed. In truth, the greatest limitation in our treatment of the disks is that the large size of our sinks means that we do not fully capture the behaviour of the inner regions of the disk. As such, there would probably be more fragmentation in the disks in reality than we find in our simulation. For a fuller discussion of the physics of disk fragmentation around population III protostars see \citep[][b]{Clark11b} where this issue is considered in great detail.

\begin{figure}
\includegraphics[width=3.5in]{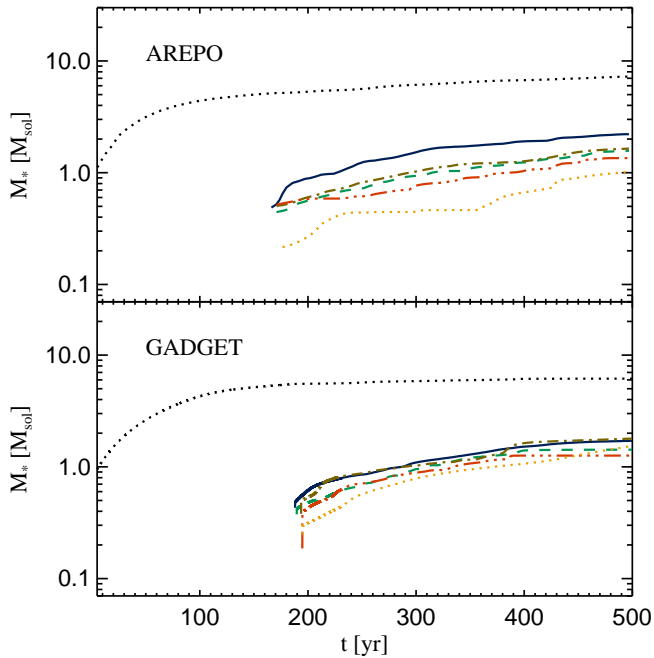}
\caption{The time of formation of sink particles and their growth in mass as measured in halo 4 without feedback. Each line represents the growth of an individual sink. The lower panel shows the standard Gadget simulation used here and the upper panel shows a comparison simulation using AREPO with the same sink properties. The method of simulation makes no significant difference to the number of fragments formed and their growth in mass in the overlapping time period.}
\label{arepo}
\end{figure}

The fragmentation here differs slightly from that shown in \citet{Greif11}, as in that case the sink radius was similar to the stellar radius, whereas here it is $20$ AU. The simulations of \citet{Greif11} used the AREPO method which has less artificial viscosity than Gadget. In order to test whether the results here were reproducible with AREPO, we re-simulated halo 4 (which showed the maximum difference between the feedback and no feedback cases) using AREPO with no feedback and the same sink sizes used in this study. \fig \ref{arepo} shows the growth of fragments formed in both simulations at the overlapping times. There is a similar interval in both cases between the first sink forming and the first burst of fragmentation. In both cases the same number of sinks form, although there are slight differences in the mass growth rates due to the different N-body dynamics which occur in each simulation. As our results are reproduced by two highly complementary numerical schemes, we are confident that we capture the true physical evolution and are not strongly influenced by numerical artefacts.

The fact that larger sinks are used here compared to the original \citet{Greif11} simulations mean that we are not resolving tight binaries and missing some young low-mass objects formed within this radius that may have been ejected. Therefore our $20$ AU sinks are a conservative estimate of the level of fragmentation. However, at this radius we are avoiding many of the uncertainties associated with protostellar mergers. As our young protostars would actually be puffy extended objects with radii about $100$ \rsun \xspace \citep{Stahler86a}, there will be strong tidal forces evoked during close interactions, leading to the possibility that fragments formed close to each other will merge when they interact. It is still unclear how best to treat this possibility. By not forming low-mass objects in close proximity to existing sinks, encounters that are close enough for the stellar radii to touch occur rarely compared to the original simulations (typically between 0-2 times in each halo) and we generally avoid this issue. However, despite these small differences, qualitatively the evolution of the halos is similar to that in \citet{Greif11}, with the main difference being that we follow the evolution for ten thousand years compared to the original thousand. 

\subsection{The effect of accretion luminosity}

\begin{figure*}
\begin{center}
\includegraphics[width=3in]{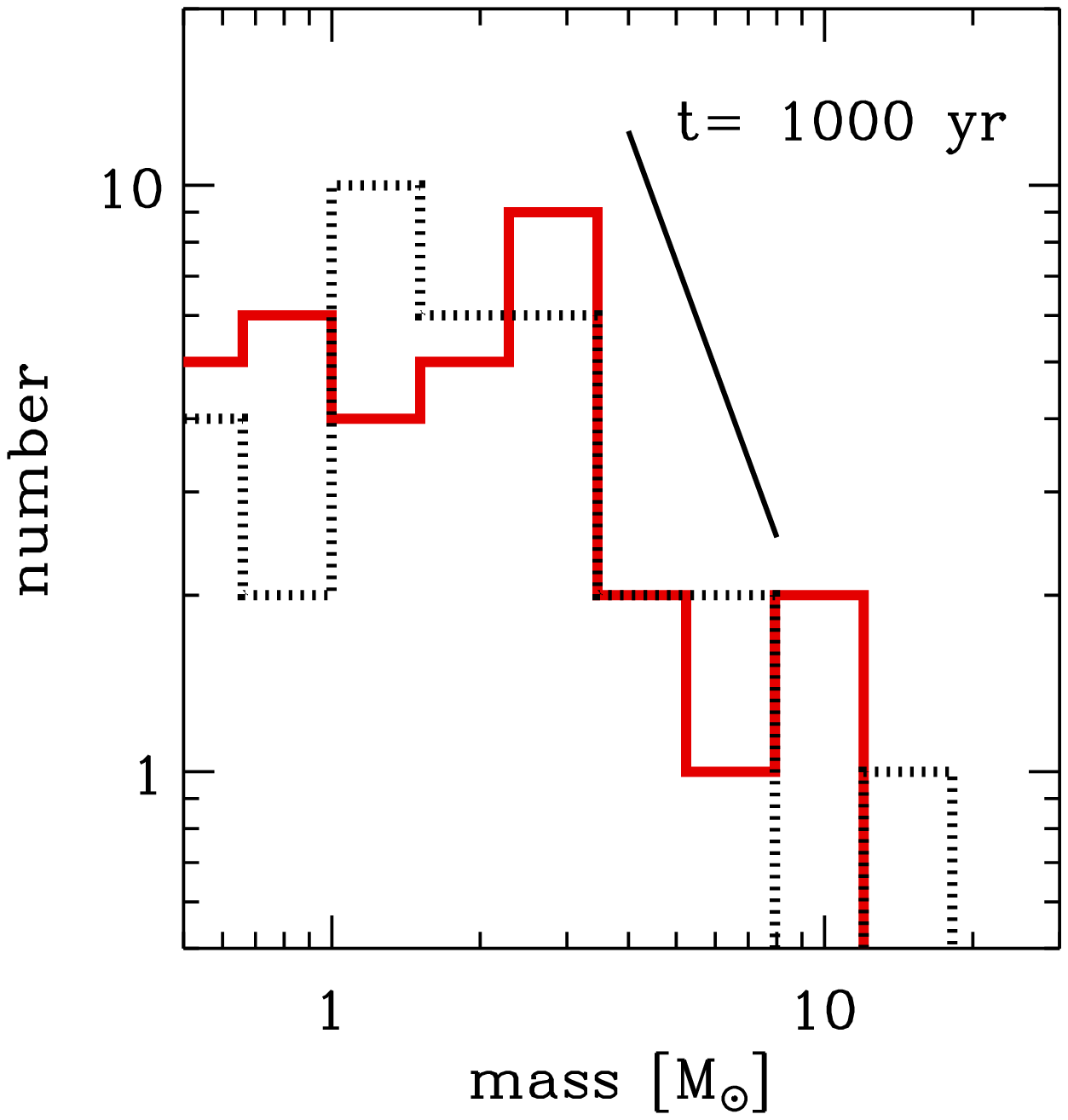}
\includegraphics[width=3in]{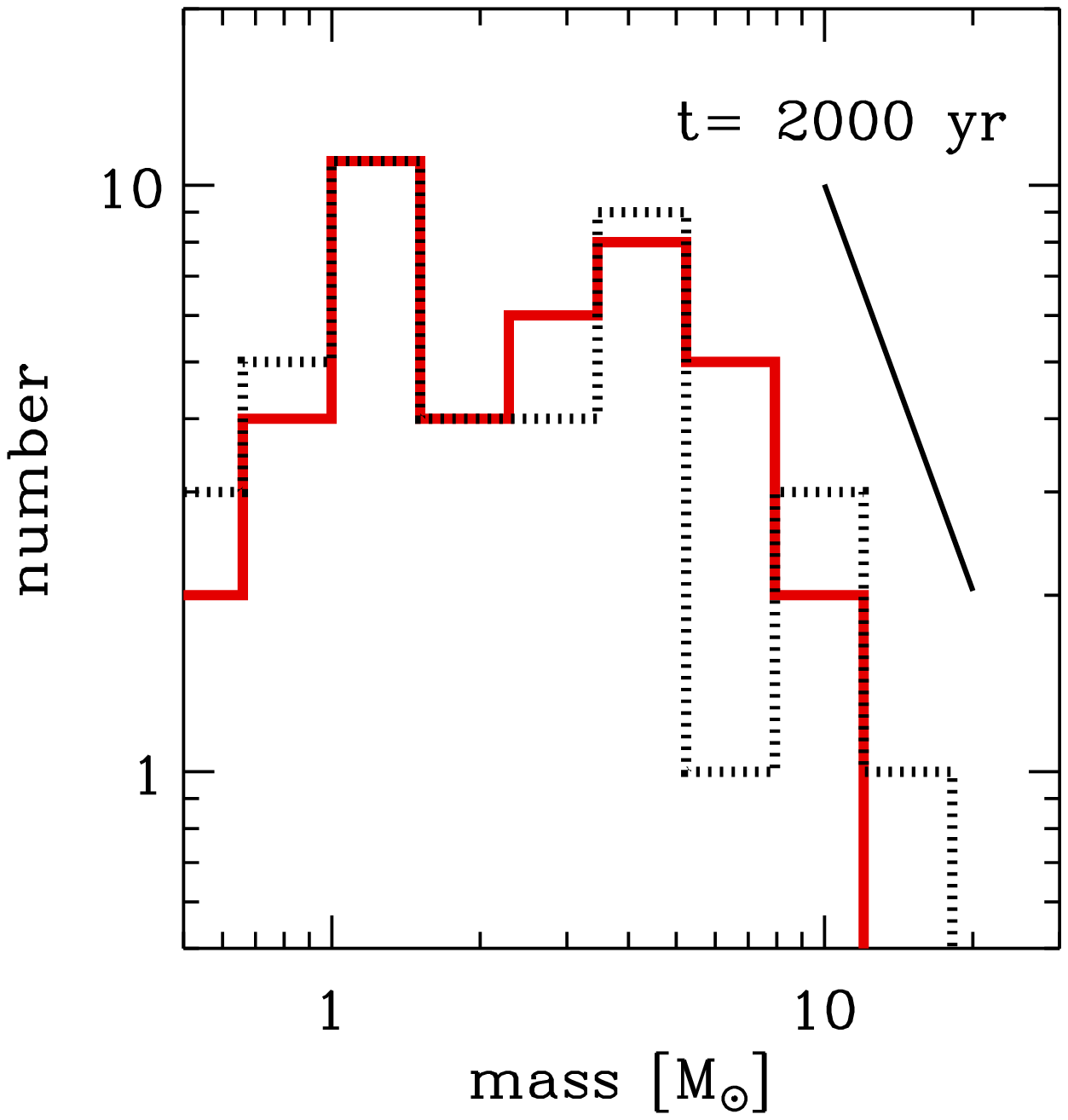}\\
\caption{The combined mass function of the sink particles formed in the five minihalos one and two thousand years after the first sink particle forms. The solid red lines shows the mass function of the halos with feedback, and the dotted black line the halos in the reference case without. In both cases the mass function is flatter than the slope of the Salpeter IMF shown by a solid black line.}
\label{mf}
\end{center}
\end{figure*}

\fig \ref{mf} shows the combined mass function of all the sinks formed in minihalos 1-5, one and two thousand years after the first sink formed. At 2000 yr in the non-feedback case ionisation effects are becoming important within Halo 5. However, for the sake of the mass function only, we run Halo 5 until this point despite the lack of ionisation in our model. This is due to the difficulty in achieving a statistically significant number of sinks for the mass function. At these early stages the sinks represent protostars rather than finished stars, and so these masses will not be those of the final population III stars. Nonetheless it can already be seen that the resulting mass function will contain a range of masses rather than just being one characteristic mass. The mass functions show no systemic variation between the case with feedback and the reference case, and both cases contain a similar total amount of mass in stars at each time. Hence the feedback has not significantly altered the fragmentation and mass growth when considering the five minihalos combined. This suggests that the results of previous studies which neglected this effect \citep[e.g.][]{Stacy10} will still be broadly correct. The mass functions appear to be flatter than the IMF's seen in the present day universe \citep{Kroupa02,Chabrier03}, although as yet we have only of order $\sim 50$ sinks, so this remains statistically uncertain.

Tables \ref{n10} and \ref{n15} show the number of fragments formed in each halo when the mass of the most massive protostar first reaches $10$ or $15$ solar masses, respectively. \citet{Tan04} find that ionising feedback does not become effective until the star is older than its Kelvin-Helmholtz time and is contracting towards the main sequence. For their fiducial model this equates to a mass of around 30\msun for a rotating protostar. However the accretion rate for the most massive object is typically only a few $10^{-3}$\myr when the protostar has reached 10\msun in our minihalos, whereas in the fiducial \citet{Tan04} models the accretion rate is $10^{-2}$\myr \xspace for a 10 \msun protostar. Since the Kelvin-Helmholtz contraction stage commences earlier with a lower accretion rate, as shown in \fig \ref{radmodel}, we estimate that ionisation feedback will become important for our minihalos when the most massive star is between $10-15$ \msun. H$_{2}$ photodissociation will also become important at this time. Beyond this point, the assumptions that we make for the luminosity heating model break down, so we chose to terminate the simulations here.

\begin{table}
	\centering
	\caption{The number of stars and time when the most massive star in the minihalo had a mass of 10\msun. The plus sign next to the number of fragments for the halo 3 reference run indicates that no star reached 10 \msun before the simulation was ended, meaning that the number of fragments in this case is a lower limit. The effects of ionising radiation are expected to become important once at least one star has reached a mass of $10 - 15$\msun or greater, and are  likely to suppress further fragmentation. There is significant inter-halo variation both in the number of fragments and the duration over which accretion luminosity is the dominant feedback mechanism. For equivalent halos, the one which forms the massive star most quickly has the least fragmentation.}
		\begin{tabular}{c c l c l }
   	         \hline
	         \hline
	         Halo & Ref. & & Feedback &  \\
	         \hline
	          & No.\ of stars & Time [yr] & No.\ of stars & Time [yr] \\
	         \hline
	         1&10 & 1,520 &10 & 2,520 \\
	         2&10 & 7,640 &7 & 4,490 \\
	         3&5+ & 9,430 &5 & 5,140 \\
	         4&17 & 7,320 &5 & 1,010 \\
	         5& 7 & 604  & 18 & 1,440 \\
	        \hline
		\end{tabular}
	\label{n10}
\end{table}

\begin{table}
	\centering
	\caption{The number of stars and time when the most massive star in the minihalo had a mass of 15 \msun. A plus sign next to the number of fragments denotes where a mass of 15 \msun was not achieved before the end of the simulation and as such the number of fragments is a lower limit. The results are similar to \tab \ref{n10} but the times are longer and there is more fragmentation.}
		\begin{tabular}{c c l c l }
   	         \hline
	         \hline
	         Halo & Ref. & & Feedback &  \\
	         \hline
	          & No.\ of stars & Time [yr] & No.\ of stars & Time [yr] \\
	         \hline
	         1&11 & 2,910 &16 & 6,040 \\
	         2&13 & 15,020 &8+ & 10,000 \\
	         3&5+ & 9,430 &6 & 11,270 \\
	         4&20+ & 22,360 &6 & 3,700 \\
	         5& 17 & 1,060  & 23 & 3,900 \\
	        \hline
		\end{tabular}
	\label{n15}
\end{table}

Perhaps the most striking feature of our calculations is the amount of variability between the outcomes of the different halos, as shown in Tables \ref{n10} and \ref{n15}. Halo 5 forms a star greater than 10 \msun \xspace after only 600 yr without feedback, whereas Halo 3 takes over 10,000 yr to do so. In the time to taken to form a 10 \msun protostar, the number of fragments varied between 5 and 17 between the halos. Therefore inter-halo variability is at least as important an effect as accretion luminosity feedback. The variability of the halos can be traced back to their chemical evolution during their collapse. \citet{Greif11} found that in two of the halos considered here (halos 2 and 3) there was significant HD cooling which allowed the gas to cool to temperatures as low as 100K \citep{Ripamonti07,McGreer08}.  When the gas was reheated by compression in the final stages of the collapse this smoothed out some of the small scale structure, resulting in less fragmentation. In the remaining three halos, HD cooling was not activated and so temperatures only as low as 200K were obtained via \h cooling. In this case, the subsequent reheating was less violent and more small scale structure was retained. A similar reduction in fragmentation has been seen in simulations of Pop III.2 star formation due to reheating \citep[][a]{Clark11a}.

\begin{figure*}
\begin{center}
\begin{tabular}{c c}
\includegraphics[width=3in]{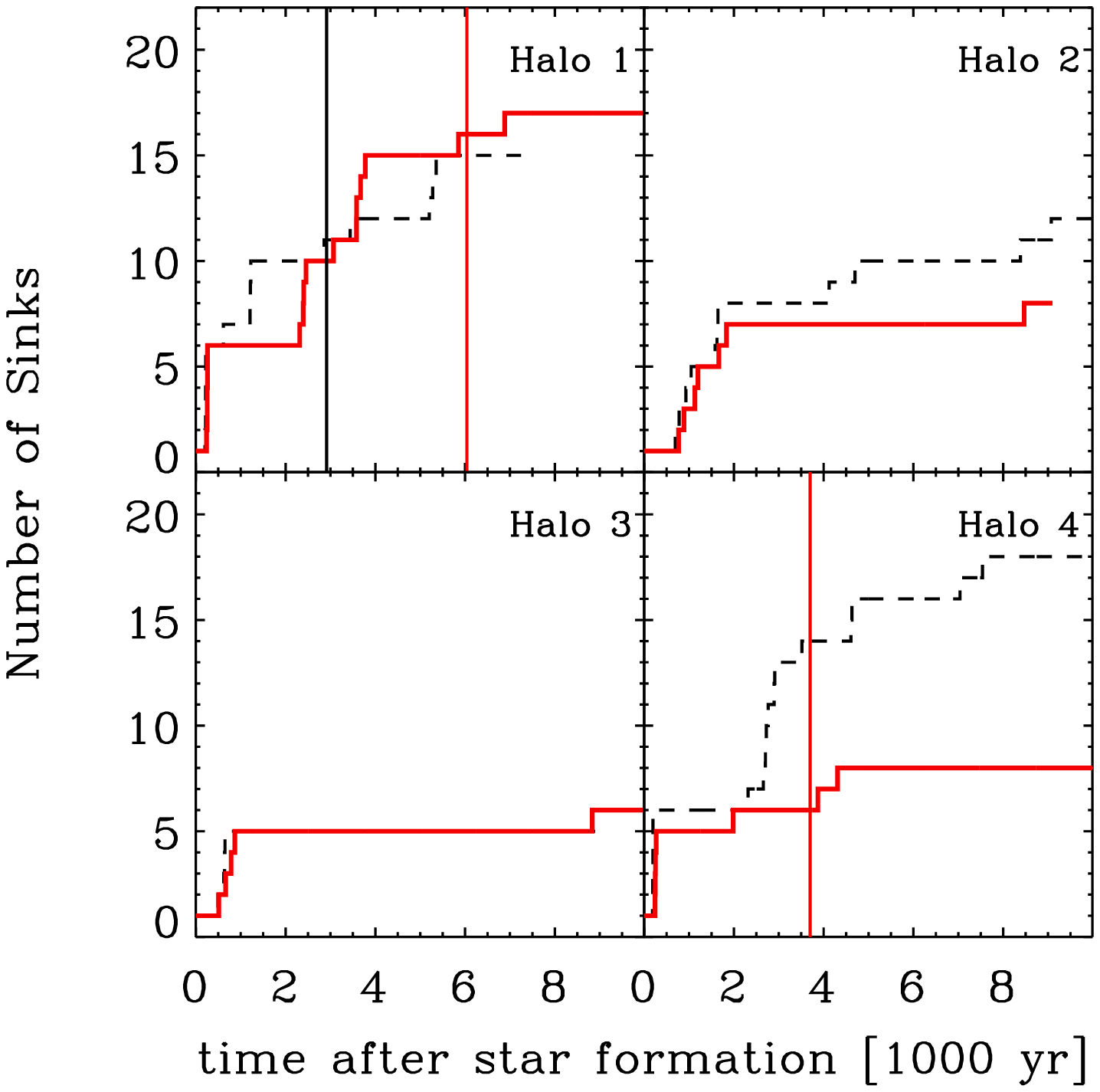}
\includegraphics[width=3in]{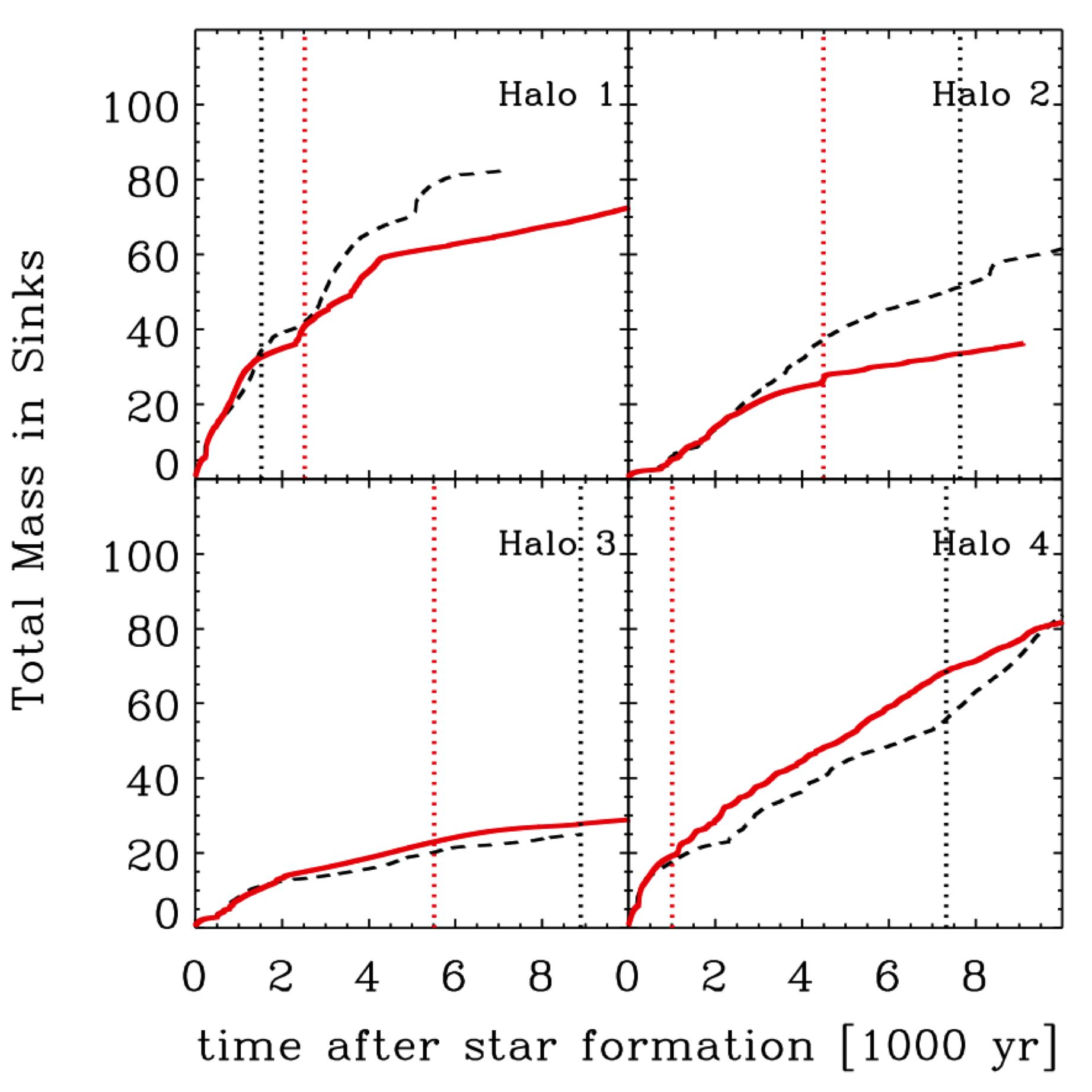}\\
\includegraphics[width=3in]{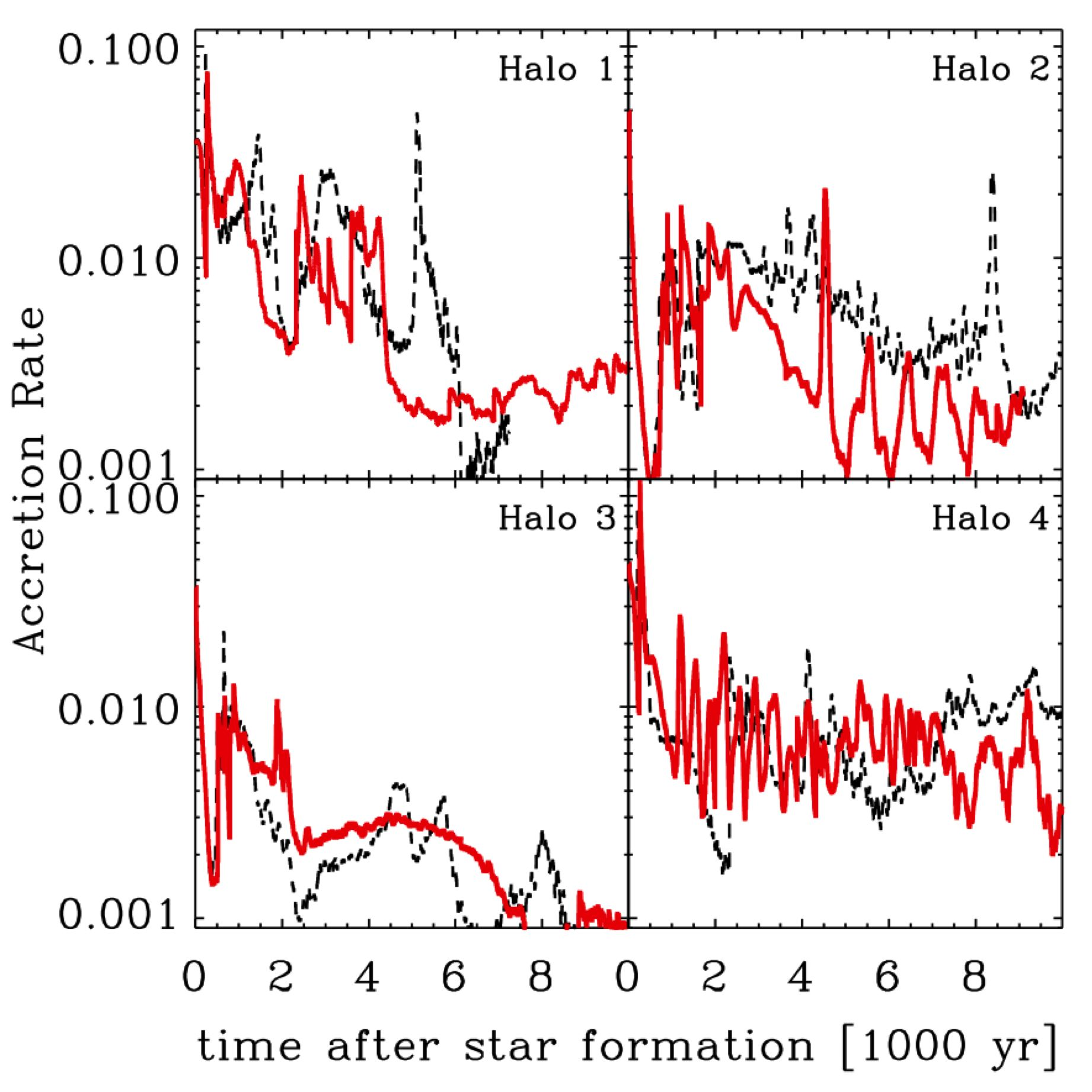}
\includegraphics[width=3in]{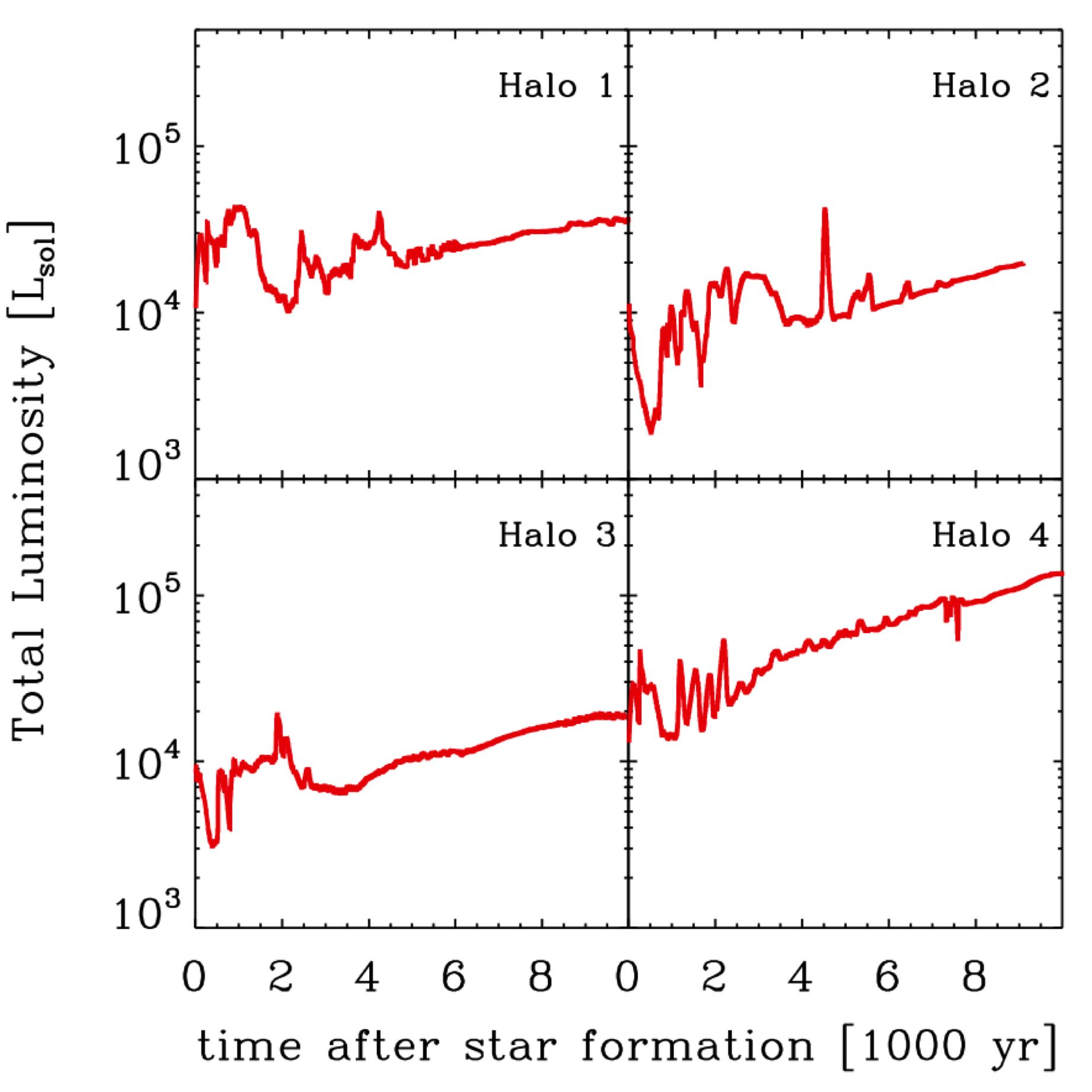}\\
\end{tabular}
\caption{The combined properties of the sinks formed in each halo plotted against the time after the first fragment forms. The solid red line shows the case with feedback and the black dotted line the case without. The vertical solid red and black lines in panel one show the point at which a star reaches 15 \msun and the dotted vertical lines in panel two show when 10 \msun has been reached.}
\label{global}
\end{center}
\end{figure*}

The first panel of \fig \ref{global} shows the evolution of fragmentation within the halos. Halos 1 and 5 fragment rapidly, meaning they quickly become dominated by chaotic dynamical interactions and the reference and feedback cases are no longer equivalent. Dynamical interactions are therefore as important as accretion luminosity effects in halos that fragment rapidly. For example, in the run of Halo 1 with feedback, there was a dynamical interaction which ejected the most massive star before it could reach 10-15 \msun, and consequently there was more time for fragmentation until one of the originally lower mass objects reached this mass. Such are the numbers of sinks formed within Halo 5 that chaotic N-body interactions cause the feedback and reference cases to swiftly diverge. Consequently no clear conclusions can be made about the effect of feedback in Halo 5 and its evolution is not shown in \fig \ref{global}.

Halos 2 and 3 are more straightforward as these halos fragment and accrete material less vigorously. As fewer fragments are formed, there is less competition to accrete the gas, which allows the first fragments to grow in mass and substantially heat their surroundings. This delays when the fragmentation occurs in the feedback case compared to the reference case. In \fig \ref{global}, fragmentation in the feedback case generally lags behind that in the reference case, and in some cases the delay can be as great as a thousand years. This was also true in Halo 1 until the chaotic dynamics made the runs diverge after a thousand years or so. The delaying of fragmentation seems to be the major consequence of accretion luminosity feedback. This was also the conclusion reached in the protostellar disk study of \citet[][b]{Clark11b}. Although feedback does not prevent fragmentation, the delay means that there are fewer fragments when ionising feedback becomes effective, so in total, accretion feedback has reduced the number of protostars formed.

Halo 4 is the case in which the maximal effect of the feedback was seen. As in Halos 1 and 5, HD cooling was not activated and 17 protostars were formed in the reference case. However, with feedback the number of fragments formed before one of the protostars reached 10 \msun was reduced from 17 fragments to 7. In Halo 4 there was enough delay before the second bout of fragmentation, after the disk first became gravitationally unstable, that the sinks were massive enough to produce a large amount of luminosity. This effect was enhanced by the geometry of the resulting system (as seen in \fig \ref{column}) with all the sink particles remaining within the central disk-like region where they could heat the dense gas.

Panel two of \fig \ref{global} shows that the total mass that goes into sink particles shows no clear correlation with feedback. Panel three of \fig \ref{global} shows that the total accretion rates have considerable temporal variation. This is firstly due to the fact that every time a new sink is formed it rapidly swallows up the gas that is bound to it, adding a large contribution to the total accretion rate. Once this gas has been accreted there is less available for the other sinks and the accretion rate can fall, a process which we term fragmentation-induced starvation \citep[][c]{Peters10c}. Another contributing factor to the total accretion rate variability is the effect of sink interactions, which we will discuss more fully in section \ref{sec:accretion}. In both the feedback and reference cases the total accretion rates are broadly similar and have a value in the vicinity of $10^{-2}$\myr \xspace initially, decreasing thereafter. The fourth panel of \fig \ref{global} shows the total luminosity of the sinks as a function of time. The total luminosity output of the stars has a value of over $10^4$ \lsun \xspace for the vast majority of the time. Fragmentation was suppressed most effectively in Halo 4 and the luminosity shown in \fig \ref{global} shows that this was indeed the case in which the feedback was most significant.

\begin{figure}
\begin{center}
\includegraphics[width=3in]{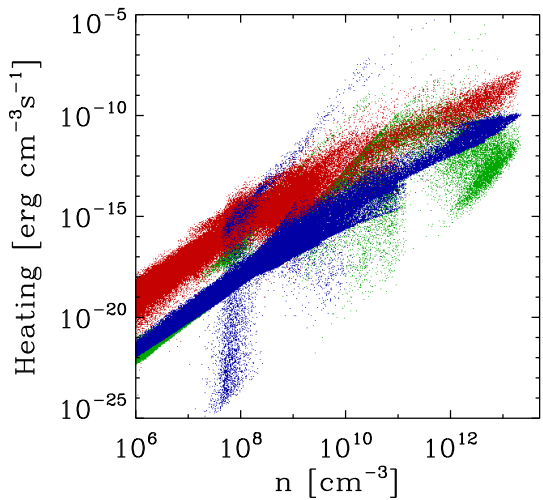}\\
\caption{The gas heating rates in Halo 1 at a typical snapshot. Red shows compressive heating, green shows heating due to \h formation, and blue shows the heating due to accretion. The heating rate from accretion is of the same order as that from \h formation, and about two orders of magnitude lower than that from compression during the collapse.}
\label{heating}
\end{center}
\end{figure}

\begin{figure}
\begin{center}
\includegraphics[width=3in]{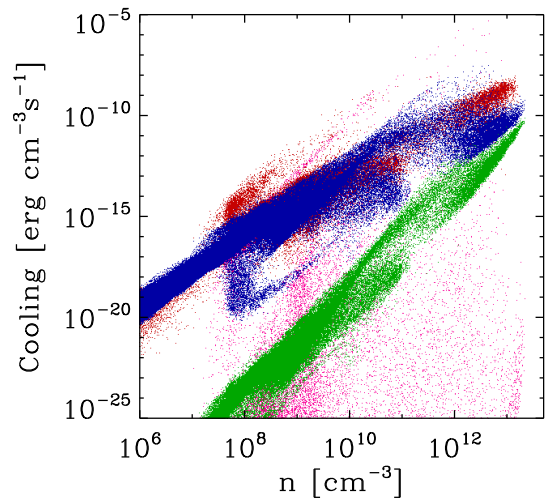}\\
\caption{The gas cooling rates in Halo 1 at the same time as \fig \ref{heating}. Blue shows \h line cooling, green shows cooling due to collisionaly induced emission, red shows cooling from re-expansion of previously compressed gas, and pink shows the cooling due to \h dissociation.}
\label{cooling}
\end{center}
\end{figure}

To understand why the accretion luminosity was unable to prevent fragmentation entirely, we need to consider the chemistry and thermodynamics of the gas more fully. \fig \ref{heating} shows the heating rates experienced by the gas from the various heat sources in Halo 1 in which feedback was largely ineffective. The heating rate from accretion luminosity is two orders of magnitude less than that from compression, and about equal to that from \h formation. We calculate the heating rate from compressional heating using the below formula that can be derived from energy conservation
\begin{equation}
\frac{d\epsilon}{dt}=-\epsilon \gamma \mathbf{\nabla \cdot v}
\end{equation}
where $\epsilon$ is the thermal energy per unit volume, $\gamma$ is the adiabatic index of the gas and v is the velocity. The heating from compression and \h formation was already being balanced by cooling from \h line emission, and re-expansion of the gas, as shown in \fig \ref{cooling}. Therefore, the addition of accretion luminosity feedback represents only a small change in the thermodynamic equilibrium of the halo. However, an important qualification is that luminosity feedback is an effect that varies with position, i.e. it is most effective close to the sinks where the gas is densest and fragmentation occurs. Additionally, the extra heating increases the collisional dissociation rate of \h, making it harder for the dense gas to cool. Consequently, the accretion luminosity heating is more dynamically significant than a first glance at \fig \ref{heating} would suggest, which is why it was able to delay fragmentation in the minihalos.

\subsection{Accretion and Dynamics}\label{sec:accretion}

The previous section demonstrates that accretion luminosity can affect the fragmentation seen in the minihalos. However it has also shown that these effects can be completely masked by the dynamics of the gas. A similar conclusion is derived from massive star formation calculations in the present day, where dynamical effects dominate over radiative feedback \citep[][a,b]{Krumholz09,Peters10b}. To explore this more fully, in this section we contrast dynamical effects with those of accretion luminosity upon the individual protostars, and consider how this will affect our feedback model.

\begin{figure*}
\begin{center}
\begin{tabular}{c}
\includegraphics[width=7in]{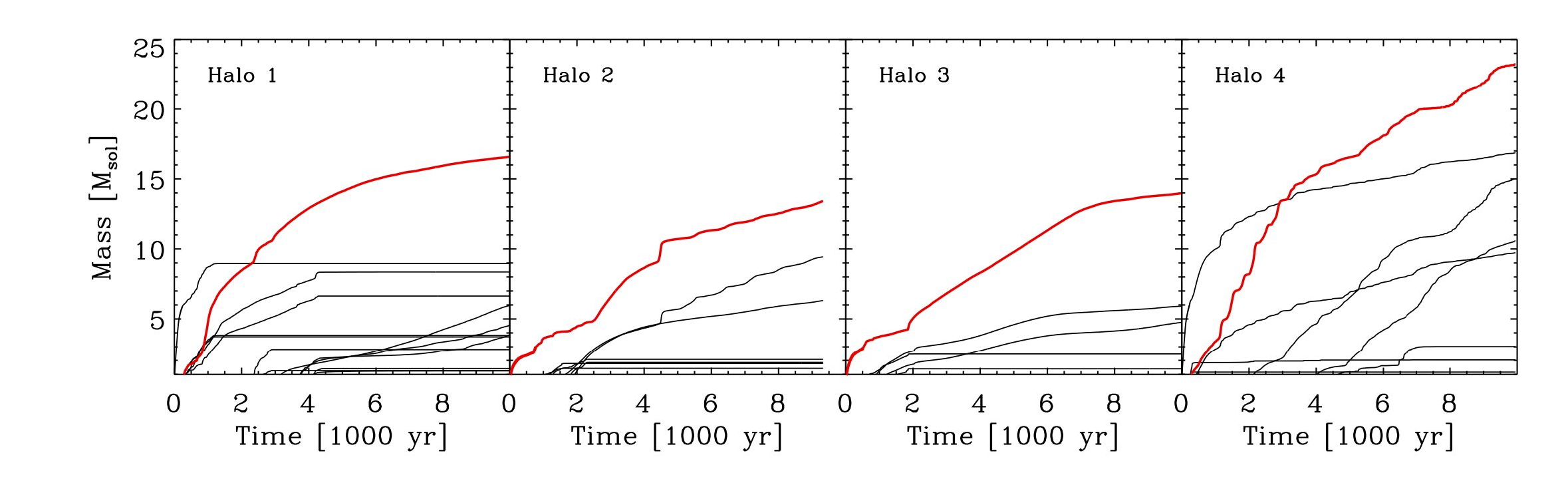}\\
\includegraphics[width=7in]{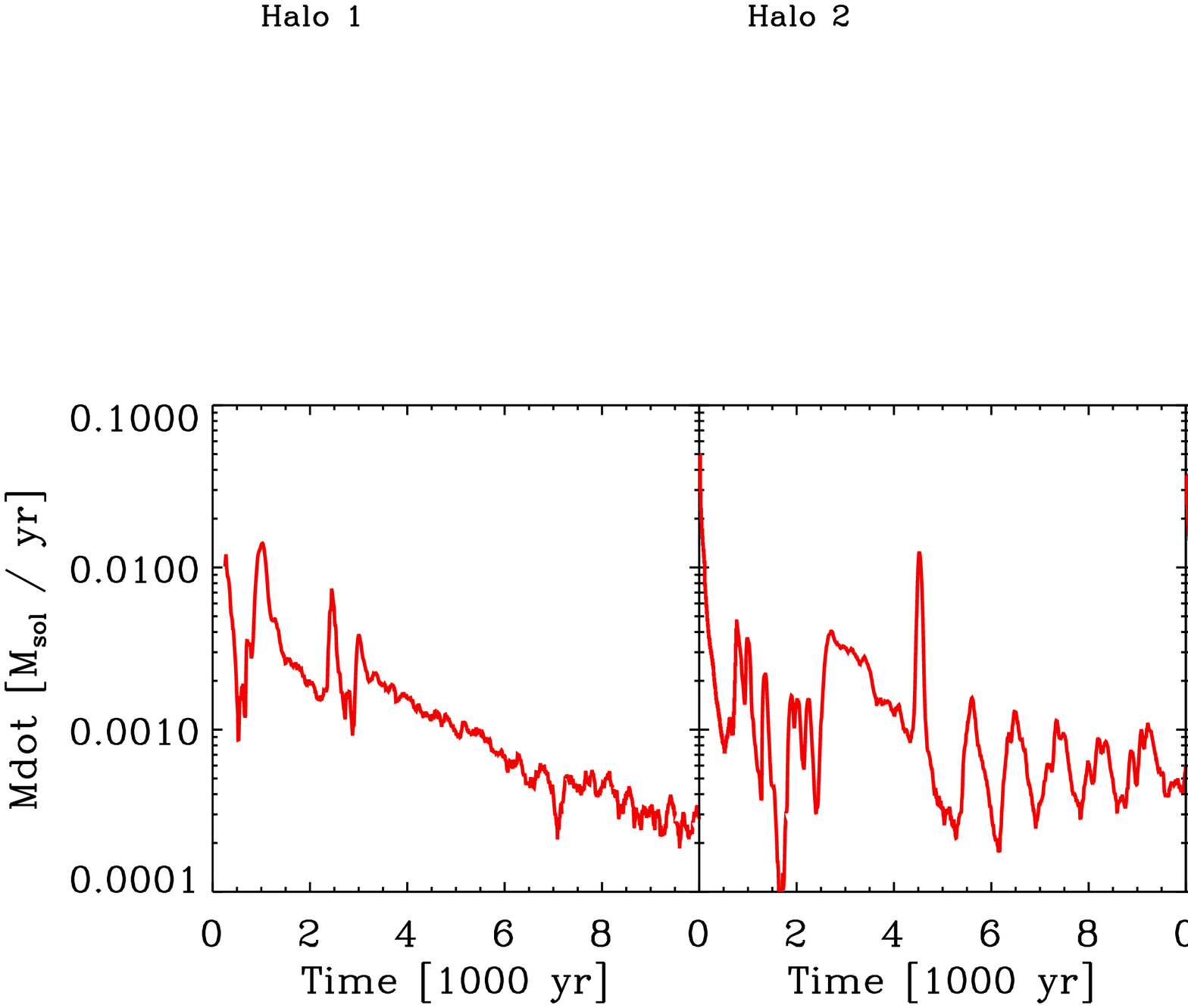}\\
\includegraphics[width=7in]{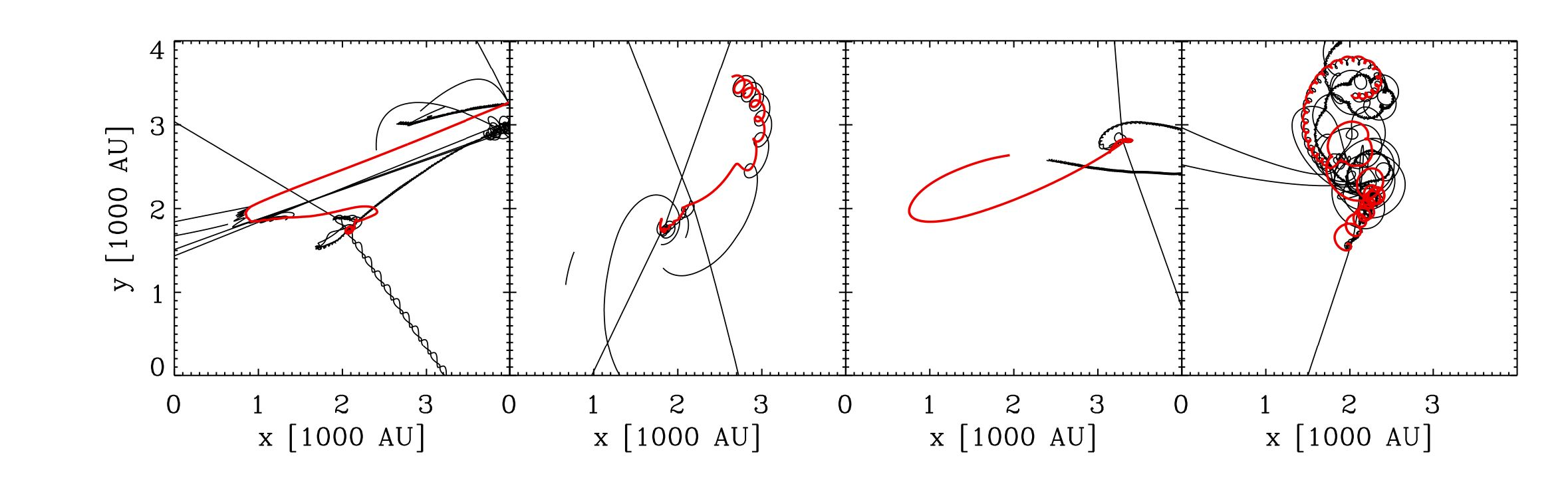}\\
\end{tabular}
\caption{The evolution of the sink particles in halos 1--4 with feedback. The most massive sink is shown in red. The accretion rates onto the sink particles are highly variable due to N-body dynamics causing sinks to orbit through their accreting gas streams.}
\label{sinks}
\end{center}
\end{figure*}

\fig \ref{sinks} shows the evolution of the sinks formed in halos 1-4. The first sink forms at the centre of the disk and quickly grows in mass with a smoothly decreasing accretion rate. \citet{Yoshida06} showed that the expected accretion rates should be as high as $10^{-1}$\myr \xspace after the first half solar mass has collapsed, and then smoothly decrease to a value of of $10^{-3}$\myr \xspace when $100$ \msun \xspace has collapsed. Our accretion rate for the initial sink agrees with \citet{Yoshida06} until fragmentation sets in. At this point the accretion rate may briefly rise as the portion of the disk between the original sink and the new sinks is strongly torqued, resulting in a large outward transfer of angular momentum and inflow of gas onto the central sink. After this short transient, the accretion rate decreases as the mass available for accretion is now shared between multiple sinks.

Once multiple sinks are formed, the accretion rates of the sink particles become highly variable. Due to the high densities characteristic of primordial star formation, the Jeans length is extremely short, and therefore fragments are formed close to each other, leading to interactions on timescales comparable to the local free-fall time. \fig \ref{sinks} also shows the paths of the stars in the central $4000$ AU during the period studied here. The sinks that remain in the halo centre orbit each other, leading to a periodically varying accretion rate as they move around in the gas bound to them. Moreover, ejections are common and occur in every halo, as shown by \citet{Greif11}. Surprisingly, even the originally central star can be ejected if it has a close three-body interaction as shown in Halo 1 (this only occurred in the case with feedback, which is why it was hard to compare Halo 1 with and without feedback). Indeed, while we find that feedback from accretion luminosity has no significant effect on the accretion rates of the protostars, dynamical interactions are extremely effective at halting an ejected protostar's accretion entirely \citep{Reipurth01,Bate02}. 

\begin{figure}
\begin{center}
\includegraphics[width=3.5in]{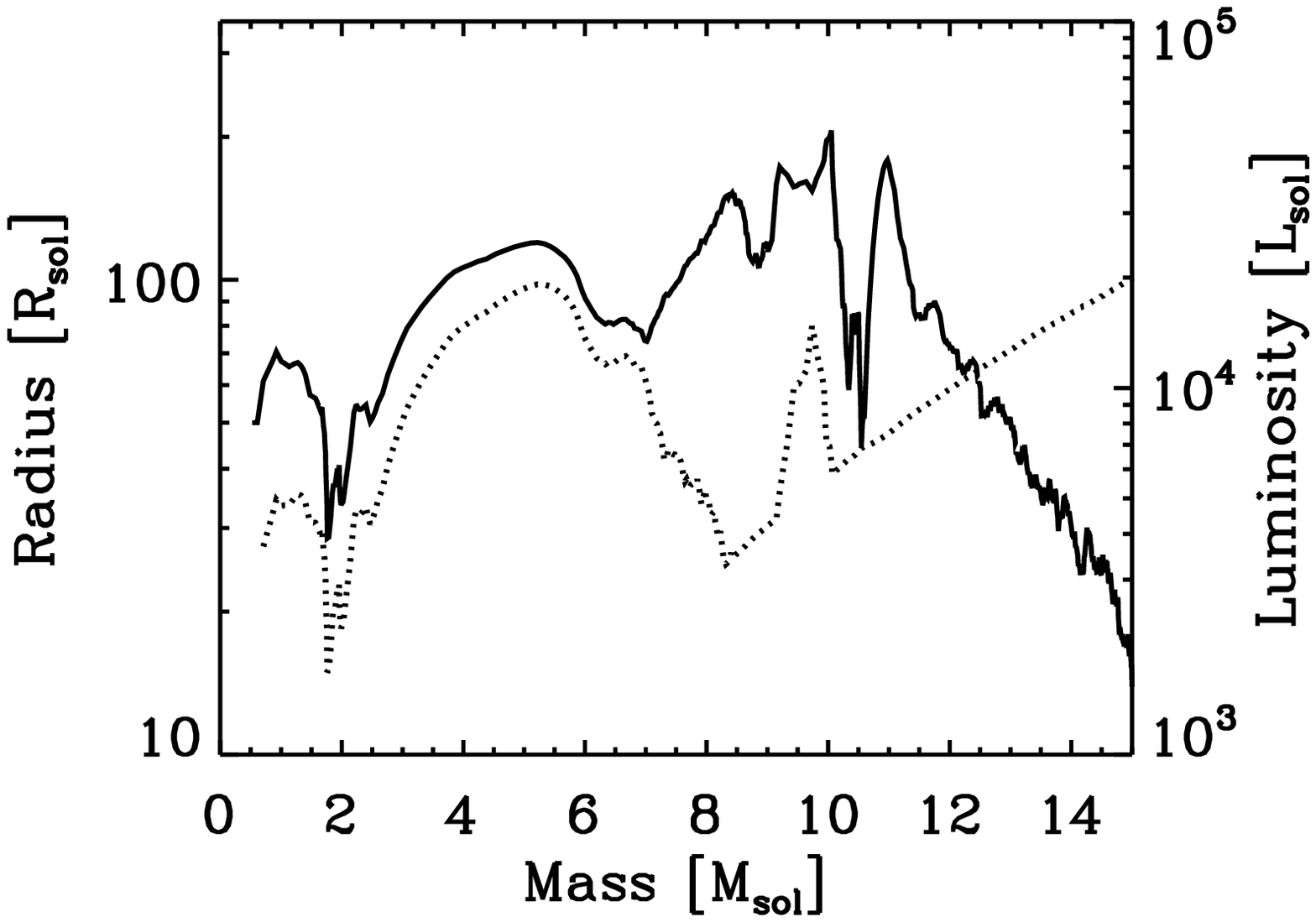}
\caption{Estimated stellar radius (solid line) and luminosity (dotted line) during the evolution of a typical massive sink.}
\label{sinkradius}
\end{center}
\end{figure}

The accretion rates for the sink particles are variable, yet the stellar radius model used to calculate the accretion luminosity was developed from simulations with a constant accretion rate.  Figure \ref{sinkradius} shows the stellar radius that results from the measured sink accretion rates using the stellar radius model shown in \fig \ref{radmodel}. The expected trend of an increasing radius which reaches a sharp peak and then rapidly decreases is still found. However, variation due to the dependency upon accretion rate is now superimposed on top of this trend. Our model is semi-analytic, not a real stellar evolution model, and therefore the sharpness in the variation is most likely artificial. In reality, the protostar would only be able to respond to changes in the accretion rate according to its Kelvin-Helmholtz time. However, allowing the radius to vary along with the accretion rate decreases the variation in the luminosity, which is proportional to $\dot{M}/R_*$, and hence this represents a conservative choice for our purposes. \fig \ref{sinkradius} also shows the accretion luminosity from the sink over the same period. Initially the variation in the accretion rate dominates both the stellar radius and the luminosity. However, over time the variation in the stellar radius becomes smaller as the star leaves the adiabatic phase and starts steadily contracting. During this later phase the actual mass of the star is no longer changing so rapidly and there is a clear trend in the radius. Consequently as \ $L_{acc}=GM_*\dot{M}/R_*$, the luminosity is no longer so noisy.

\section{Discussion}
The effect of accretion luminosity feedback is a modifying factor affecting fragmentation in minihalos, rather than a dominant one. Inter-halo variability produces a greater variation in the number of fragments formed than feedback effects, and the number of fragments chiefly depended on the initial conditions of the halo which was being considered. In halos in which a large number of fragments formed, N-body interactions and ejections are at least as important as accretion luminosity and play an important role in the accretion histories of the protostars. It was only for more slowly fragmenting halos in which only a few fragments were initially formed, that the accretion feedback became important. At this point, the chief role of the accretion feedback was to delay fragmentation. This limits the number of fragments that can form before the largest sink particle becomes massive enough to start emitting significant quantities of ionising radiation. Even with the inclusion of feedback, a small cluster of sink particles was formed in all the minihalos studied here. 

However the analysis of the accretion rates and dynamics has raised some interesting questions in its own right. The accretion rates onto the sink particles were highly variable, even periodically so in some cases. The accretion rates were recorded at the sink accretion radius of 20~AU instead of directly onto a protostar, and so some of this variation may be smoothed by an inner disk. However, it is unlikely this effect could be removed entirely. Population III stellar modeling \citep[e.g.][]{Omukai03} typically assumes a constant accretion rate, and it is unclear how a variable accretion rate would affect the stellar evolution of Population III protostars. 

Moreover, as highlighted by \citet{Greif11}, there are a large number of encounters that can produce ejections. In treating encounters we use a simple sink particle prescription that considers the stars as point masses, whereas in reality the protostars are extended gaseous objects with radii of typically around   $100$ \rsun. Tidal forces will be strong during such an encounter raising the possibility the protostars might merge. The two approaches that are typically used for close encounters with sink particles are either to model the interaction as if it were occurring between point masses (as done here) or to combine the particles together as `sticky' particles. Both approaches are probably oversimplifications of the true picture, and given that we have found that encounters are more important than feedback in determining the early history of population III protostars, it would be useful if this issue were addressed in the future.

Another consequence of this work is the implications for producing `dark stars'. It has been proposed that dark matter annihilations at the centre of a minihalo, where the dark matter density is at a maximum, could be a significant source of energy that could support primordial stars with radii of up to 10 AU \citep{Spolyar08,Spolyar09,Freese08}. However in the minihalos studied here, the protostars never precisely remain at the centre of their dark matter halos throughout this early stage of their evolution. It is, however, too early to say whether these findings exclude dark stars as the predicted dark matter annihilation luminosity is at least an order of magnitude greater than that found here from accretion feedback. \citet{Ripamonti10} found that when full gas chemistry was include in a 1D calculation, annihilation feedback was not sufficient to halt collapse and a normal population III protostar was formed. However, the contribution from annihilation feedback may be enough to prevent fragmentation, in which case the protostars would have no interactions and remain in the centre of their halos. This is a question that we are currently addressing. 

\section{Conclusions}
We have introduced a prescription for heating from accretion luminosity into a re-simulation of five minihalos from cosmological initial conditions. We followed the evolution of these halos with and without feedback up until the point at which ionisation feedback would become significant. Our findings are:\\
\begin{enumerate}
\item Accretion luminosity delays fragmentation but cannot prevent it.
\item The intrinsic variation in halo properties due to differences in their formation history generally has a larger effect on the number of fragments formed than the accretion luminosity does.
\item Halos in which a large number of fragments form rapidly are dominated by dynamical effects. It is only in more slowly fragmenting cases that form fewer fragments that accretion luminosity becomes effective.
\item Accretion luminosity has little to no effect on the accretion rates of the protostars. On the other hand,  dynamical ejections are an effective means of halting further accretion.
\item The accretion rates measured for the sink particles are highly variable and are quite different from the constant or slowly varying accretion rates assumed in most pre-main sequence models for Population III stars.
\end{enumerate}

\section*{Acknowledgements}
We would like to particularly thank Takashi Hosokawa and Kazuyuki Omukai for sharing the data used to develop our stellar radius model and for all the useful feedback they provided. We would also like to thank Dominik Schleicher, Volker Bromm, Naoki Yoshida, Tom Abel, Matt Turk, and Volker Springel for useful discussions which added to this paper. We acknowledge financial support from a number of sources: the Baden-W\"{u}rttemberg Stiftung via their program International Collaboration II (grant P-LS-SPII/18); a Frontier grant of Heidelberg University sponsored by the German Excellence Initiative; NSF grants AST-0708795 and AST-1009928; and NASA through Astrophysics Theory and Fundamental Physics Program grants NNX 08-AL43G and 09-AJ33G (V.B.). In addition, we are grateful for subsidies from the German Bundesministerium f\"{u}r Bildung und Forschung via the ASTRONET project STAR FORMAT (grant 05A09VHA) as well as from the Deutsche Forschungsgemeinschaft (DFG) under grants no.\ KL 1358/1, KL 1358/4, KL 1359/5, KL 1358/10, and KL 1358/11.

\bibliography{Bibliography}
\label{lastpage}

\end{document}